\newif\ifpnas
\pnasfalse

\ifpnas
\documentclass[9pt,onecolumn,twoside,nofootinbib]{pnas-new}
    \usepackage{dsfont}
    
\usepackage{verbatim}
\usepackage{amsmath}
\usepackage{comment}
\usepackage{amsfonts}
\usepackage{amssymb}
\usepackage{numprint}
\usepackage{bm,bbm}
\usepackage{fancyhdr}
\usepackage{empheq}
\usepackage{color} 
\else
  \documentclass[onecolumns,reprint,floatfix,nofootinbib]{revtex4-1}
      \usepackage{dsfont}
    \usepackage{comment}
  \usepackage[utf8]{inputenc}
  \usepackage{newtxtext}
  \RequirePackage[dvipsnames,usenames]{xcolor}
  \usepackage{hyperref}

  \usepackage{color} 
  \usepackage{amsmath} 
  \usepackage{graphicx}

\usepackage{cancel}
\usepackage{url}
\usepackage{xcolor}
\usepackage{bm,bbm}
\usepackage{fancyhdr}
\usepackage{empheq}

\usepackage{graphicx}
\usepackage{amsmath}
\usepackage{amsfonts}
\usepackage{amssymb}
\usepackage{numprint}

\fi

\ifpnas
    \templatetype{pnasresearcharticle} 
\else
  \makeatother
 \begin{document}
\fi

\title{Distribution of centrality measures on undirected random networks via cavity method
}
\ifpnas

\author[a,b]{Silvia Bartolucci}
\author[a,c]{Fabio Caccioli}
\author[d]{Francesco Caravelli}
\author[e]{Pierpaolo Vivo}

\affil[a]{Dept. of Computer Science, University College London, 66-72 Gower Street WC1E 6EA London (UK)}
\affil[b]{Centre for Financial Technology, Imperial College Business School, South Kensington SW7 2AZ London (UK).} 
\affil[c]{ London Mathematical Laboratory, 8 Margravine Gardens, London WC 8RH (UK).}
\affil[d]{Theoretical Division,
Los Alamos National Laboratory, Los Alamos, New Mexico 87545 (USA)}
\affil[e]{Dept. of Mathematics, King's College London, Strand WC2R 2LS London (UK).}


\leadauthor{S. Bartolucci}

\significancestatement{}

\authorcontributions{XXX}
\authordeclaration{No conflict of interest.}
\correspondingauthor{\textsuperscript{2}To whom correspondence should be addressed. E-mail: pierpaolo.vivo@kcl.ac.uk}

\keywords{centrality | large-N | low-rank approximation| degree-centrality | networks}

\else
\author{Silvia Bartolucci$^{1,2}$, Francesco Caravelli$^3$, Fabio Caccioli$^{1,4,5}$, Pierpaolo Vivo$^{6,\star}$}
\vspace{1cm}
  \affiliation{\\\vspace{0.5cm}$^1$ Dept. of Computer Science, University College London, 66-72 Gower Street WC1E 6EA London (UK).\\ $^2$Centre for Financial Technology, Imperial College Business School, South Kensington SW7 2AZ London (UK). \\ $^3$ Theoretical Division (T4), Condensed Matter \& Complex Systems, Los Alamos National Laboratory, Los Alamos, New Mexico 87545 (USA). \\ $^4$ Systemic Risk Centre, London School of Economics and Political Sciences, WC2A 2AE, London (UK).\\ $^5$ London Mathematical Laboratory, 8 Margravine Gardens, London WC 8RH (UK). \\ $^6$ Dept. of Mathematics, King's College London, Strand WC2R 2LS London (UK).\\
  $^\star$ Corresponding author: {\texttt{pierpaolo.vivo@kcl.ac.uk}}}

\fi

\begin{abstract}
The Katz centrality of a node in a complex network is a measure of the node's importance as far as the flow of information across the network is concerned. For ensembles of locally tree-like and undirected random graphs, this observable is a random variable. Its full probability distribution is of interest but difficult to handle analytically because of its ``global'' character and its definition in terms of a matrix inverse. Leveraging a fast Gaussian Belief Propagation-cavity algorithm to solve linear systems on a tree-like structure, we show that (i) the Katz centrality of a single instance can be computed recursively in a very fast way, and (ii) the probability $P(K)$ that a random node in the ensemble of undirected random graphs has centrality $K$ satisfies a set of recursive distributional equations, which can be analytically characterized and efficiently solved using a population dynamics algorithm. We test our solution on ensembles of Erd\H{o}s-R\'enyi and scale-free networks in the locally tree-like regime, with excellent agreement. The distributions display a crossover between multimodality and unimodality as the mean degree increases, where distinct peaks correspond to the contribution to the centrality coming from nodes of different degrees. We also provide an approximate formula based on a rank-$1$ projection that works well if the network is not too sparse, and we argue that an extension of our method could be efficiently extended to tackle analytical distributions of other centrality measures such as PageRank for directed networks in a transparent and user-friendly way.
\end{abstract}

\ifpnas
  \dates{This manuscript was compiled on \today}
  \doi{\url{www.pnas.org/cgi/doi/10.1073/pnas.XXXXXXXXXX}}

  \begin{document}

  \maketitle
  \thispagestyle{firststyle}
  \ifthenelse{\boolean{shortarticle}}{\ifthenelse{\boolean{singlecolumn}}{\abscontentformatted}{\abscontent}}{}




\else
  \maketitle

\fi

\ifpnas  \else 

\section{Introduction}

The study of complex systems as well as the applications of the ``science of complexity'' to the most disparate areas of research have witnessed spectacular successes in recent years. 
Complex systems are quintessentially defined as being composed of many components that are interacting locally, exhibiting emerging static and dynamical properties, and involving a certain degree of randomness. However, not every elementary constituent plays the same role in the structure or functionality of a system, with some constituents being more critical and ``central'' to ensure stability, resilience, or other desired global properties of the architecture \cite{Albert2000,Callaway2000,jeong2001lethality,de2014navigability,boccaletti2006complex, Goltsev2012,Gao2016, CruaAsensio2017,Farooq2019,Guilbeault2021,epidemiology,bbv-book,protein,battiston2012debtrank,bardoscia2021physics,intro1,barabasirank}. 
Identifying the most important nodes in a network architecture is indeed of paramount importance to ensure the integrity and functionality of transportation networks and critical infrastructures \cite{Guimera2005, Wu2006, brown2006defending, Carvalho2009, Duan2014}, as well as to allow users to retrieve an accurate list of webpages corresponding to an Internet query \cite{Prank,page1999pagerank}, or identify the most suitable receivers of a vaccine to mitigate a disease outbreak \cite{Kitsak2010,Salathe2010,Wang2016,Pung2022}. Our ability to exploit the advantages of living in a modern and interconnected society to the full heavily relies on preserving the integrity of crucial infrastructure such as the Internet and power grids \cite{Albert2000, Sole2008, doyle2005robust,rinaldi2001identifying, Cohen20001, Schafer2018}.  

Several ``centrality'' measures have been devised to classify and rank nodes of a network, which focus on different structural characteristics: the \emph{degree centrality} simply counts how many neighbors each node has and ranks nodes according to how \emph{locally} connected they are. More global centrality measures include the eigenvector centrality \cite{bonacich1972factoring}, the Katz centrality mainly considered here \cite{katz1953new}, and Google PageRank \cite{Prank,page1999pagerank}. Other definitions take into account the relative position of each node in the network (for instance, closeness and betweenness \cite{freeman1977set,corr1}, communicability \cite{estradabook,comm4,comm2,comm3} and DomiRank \cite{domirank}), as well as the role played by a node in a dynamic process on networks (for instance, current-flow \cite{brandes2005centrality}, entanglement \cite{ghavasieh2021unraveling}, and random-walk \cite{PhysRevE.90.032812} centralities) -- see \cite{generalcentrality} and references therein for a taxonomy of centrality measures on networks and \cite{reviewcentrality,reviewcentrality2,reviewcentrality3} for comprehensive reviews. 

When the underlying structure is a single instance or an ensemble of \emph{random} networks, generated according to probabilistic rules, each of the above centrality measures becomes a random variable, whose precise statistics is of general interest. Indeed, distributions of observables on random graphs constitute an important benchmark, as ``null models'' constructed out of random interactions can then be compared with empirical data to quantify the effect of structure and ``information'' encoded in the data that cannot be explained by pure noise.

Perhaps surprisingly, though, the available analytical results for the full distribution of centrality measures on random networks are particularly scarce. This is probably due to the ``global'' character of most centrality measures, which require the full and complete information about all other nodes to be characterized exactly.

In the recent mathematical literature, most of the existing works concern the distribution of PageRank on \emph{directed} random graphs \cite{pagerankdistribution1,pagerankdistribution2,pagerankdistribution3,pagerankdistribution4,pagerankdistribution5,pagerankdistribution6,pagerankdistribution7,pagerankdistribution8,pagerankdistribution9}, in particular, aimed at proving rigorously the empirically observed `power-law hypothesis': in a scale-free network, the PageRank scores follow a power law with the same exponent as the (in-)degrees \cite{powerlawSF1,powerlawSF2,powerlawSF3,powerlawSF4,powerlawSF5}. In this context, the distribution of PageRank was found to obey a distributional fixed-point equation, which seemingly facilitated analytical considerations. However, the derivations are not particularly transparent or illuminating -- at least to our eyes -- and do not allow easy access to an operational scheme to control and solve the distributional equations. Upper bounds and approximations to the PageRank distribution are provided in \cite{uppSF} for $d$-regular directed acyclic random networks generated by the configuration model. The distribution of betweenness centrality was considered for exponential random graph models in \cite{duron} and for random trees and other subcritical graph families in \cite{durant}. Exact calculations of centrality vectors for instances of networks with special structures are also available \cite{paton}. For undirected random graphs, bounds and convergence of the PageRank distribution have been obtained in \cite{undirected1}, while numerical explorations of distributions of various centrality measures (including PageRank) as well as analytical results for networks with preferential attachment are presented in \cite{perra}. For an empirical study of the distribution of centralities in urban settings, see \cite{latoraurban,urban2}.

In this paper, we focus on the Katz centrality of undirected random networks with $N$ nodes that are \emph{locally tree-like}, meaning that short loops are rare and the typical size of a loop is $\mathcal{O}(\log N)$. However, our techniques work also in the case of other similarly constructed centrality measures \cite{bart1}. We aim to characterize analytically the full distribution of the Katz centrality of nodes (i) within a single instance with $N$ nodes, and (ii) across the entire ensemble of large random graphs with fixed mean degree $c$ for $N\to\infty$, focusing on Erd\H os-R\'enyi and Scale Free graphs as prominent examples\footnote{While power-law networks with exponent less than $3$ have finite loops \cite{SFloops}, the tree-like approximation appears to work well also on these structures \cite{Goltsev2012}.} -- although the theory works as well for any configuration model characterized by the degree distribution $p(k)$.  

Leveraging a fast recursive scheme based on cavity/Gaussian Belief Propagation (GaBP) to solve linear systems on a tree-like structure \cite{GaussianBP,linearshental,linsyst}, we first show that the Katz centralities of all nodes of a single instance solve a system of recursive equations for cavity fields, which can be solved very efficiently. Next, we exploit this result to claim that the corresponding distribution of Katz centralities across the entire ensemble can be determined as the solution of a set of recursive distributional equations -- essentially, integral equations for probability density functions (pdf). Not only are these equations written out explicitly, but an efficient numerical scheme (Population Dynamics) is proposed to solve them numerically, the only necessary ingredient being the degree distribution $p(k)$ of the network of interest. The numerical solution of the population dynamics scheme is in excellent agreement with numerical simulations of large random networks with fixed average connectivity.

We also propose an approximate scheme -- based on a rank-$1$ projection of the adjacency matrix proposed in \cite{bart1} and successfully used in \cite{bart2,bart3} -- to reproduce the distribution of Katz centrality for not too sparse graphs, which also works very well. All our results confirm and put on firmer analytical ground the known observations that centrality measures are often correlated with each other \cite{evans,corr2,corr3,corr4}, as we show that the distribution of Katz centrality can be naturally decomposed into contributions coming from nodes of given degree (see Eq. \eqref{PKsuperposition} below) yielding a strong correlation between Katz and degree centrality of each node (see Fig. \ref{fig:single} and \ref{fig:single2} below).

We will also argue that an extension of our framework is likely to be useful to compute analytically the full distribution of other centrality measures (for example, PageRank in \emph{directed} graphs) in a transparent and easy-to-interpret way.

The plan of the paper is as follows. In Section \ref{sec:Katz} we provide the definition and interpretation of Katz centrality, and we show that the centralities of nodes can be computed as the solution of a linear system. In Section \ref{sec:linearsys} we provide a pedagogical derivation of the cavity/BP recursive equations that allow us to solve a sparse linear system of equations on a tree-like structure in a fast and efficient way. In Section \ref{sec:KatzSI} we leverage this result to derive a set of recursive equations to compute the Katz centrality of all nodes of a single instance of a network in a fast and distributed way. In Sec. \ref{sec:PK} we exploit these results to show that the full probability distribution $P(K)$ of observing a node with Katz centrality $K$ in an ensemble of large random networks is determined as the solution of a pair of recursive distributional equations, which can be efficiently solved using a Population Dynamics algorithm presented in Sec. \ref{sec:popdyn} along with the result of numerical simulations. In Section \ref{sec:rank1} we construct an approximate scheme -- based on a rank-$1$ projection of the adjacency matrix -- to write $P(K)$ in a more explicit form, which works well in certain conditions. Finally, in Section \ref{sec:concl} we offer some concluding remarks and an outlook for future research.

\section{Katz centrality}\label{sec:Katz}

 In graph theory, the Katz centrality of a node was first introduced by Leo Katz in 1953 \cite{katz1953new} to measure the relative degree of influence of an agent within a social network by taking into account the total number of walks that connect the agent with all the others. Paths connecting an agent with a ``distant'' node are however penalized by an attenuation factor $\alpha$.

More formally, let $G$ be the $N\times N$ symmetric adjacency matrix of an undirected network formed by $N$ nodes, with $G_{ij}=G_{ij}=1$ if node $i$ is connected to node $j$, and $0$ otherwise. The powers of $G$ indicate the presence (or absence) of links between two nodes through intermediaries. For instance, the element $(G^k)_{ij}$ indicates that there is a path of length $k$ between nodes $i$ and $j$.

Given a parameter $\alpha\in (0,1)$, $K_i$ denotes the Katz centrality of node $i$ if
\begin{equation}
    K_i=\sum_{k=1}^\infty \sum_{j=1}^N \alpha^k (G^k)_{ji}\ .\label{eq:centrsum}
\end{equation}
The interpretation is clear: the centrality of a node is a weighted sum of paths of all lengths reaching that node from all other nodes, where longer paths are weighted less -- see \cite{matching} for proposals on how to optimally select the parameter $\alpha$.

The value of the attenuation factor $\alpha$ has to be chosen such that 
\begin{equation}
    0<\alpha<\frac{1}{\lambda_{max}}\ ,\label{alphalambda}
\end{equation}
where $\lambda_{max}$ is the largest eigenvalue of $G$, for the infinite sum in \eqref{eq:centrsum} to converge. Interestingly, it follows from the definition in \eqref{eq:centrsum} that 
\begin{equation}
    \lim_{\alpha\to 0^+}\frac{K_i}{\alpha}=k_i\ ,
\end{equation}
where $k_i = \sum_j G_{ji}$ is the degree of node $i$, i.e. the number of its neighbors. Conversely,
\begin{equation}
    \lim_{\alpha\to (1/\lambda_{max})^-}(1-\alpha\lambda_{max}) K_i=\xi E_i\ ,
\end{equation}
where $E_i$ is the eigenvector centrality of node $i$, i.e. the $i$-th component of the vector $\bm E$ that solves the eigenvector equation $G\bm E = \lambda_{max}\bm E$, and $\xi$ is a numerical constant, see e.g. \cite{EV}.

 The infinite geometric sum in \eqref{eq:centrsum} converges to
 \begin{equation}
     \bm K = \underbrace{(\mathds{1}-\alpha G)^{-1}\bm 1}_{\bm K_s}-\bm 1\ ,\label{defKatzvector}
 \end{equation}
 where $\mathds{1}$ is the $N\times N$ identity matrix, and $\bm 1$ is a $N$-dimensional column vector. Here, $\bm K$ is the vector collecting the $N$ centralities of all nodes. From \eqref{defKatzvector} and the fact that $\alpha G$ is sub-stochastic, it follows\footnote{We have $(\bm K_s)_i\geq 0$ from \cite{KSgreater1}. Then, $(\bm K_s)_i=1+\alpha (G\bm K_s)_i$ from \eqref{defKs}. Since $G$ has non-negative entries and $\alpha$ is non-negative, the claim easily follows.} that $K_i\geq 0$.

 Rearranging Eq. \eqref{defKatzvector} slightly, we can rewrite the vector of centralities as the solution of the linear system of equations
 \begin{equation}
    (\mathds{1}-\alpha G)\bm K_s =\bm 1\ ,\label{defKs}
 \end{equation}
where $\bm K_s=\bm K + \bm 1$.

In the following section, we review the algorithm to solve efficiently a linear system of equations on a sparse structure using a recursive method (GaBP/cavity) \cite{GaussianBP,linearshental,linsyst}, and then we apply it to the linear system at hand. Standard iterative schemes for linear systems such as Gauss-Seidel, Jacobi, and conjugate gradient \cite{iterative} are routinely used to numerically compute the centrality values on a single instance \cite{conjugategradient}, as they are more stable and faster than matrix inversion methods. The GaBP/cavity scheme we propose to employ here has however two main advantages: (i) there is some numerical evidence that the GaBP/cavity scheme is superior to standard recursive linear system methods in terms of performances and stability on sparse structures \cite{performanceBP,performanceBP2}, and (ii) contrary to classical recursive method, the GaBP/cavity scheme provides explicit equations connecting single-instance node and edge fields, which can be easily translated into analytical distributional equations at the ensemble level. We start in the next section by presenting the general GaBP/cavity theory for the solution of sparse linear systems.

\section{Solution of a sparse Linear System with cavity method}\label{sec:linearsys}

Consider a linear system
\begin{equation}
    A\bm x=\bm b \label{linearsys}
\end{equation}
with $A$ square, symmetric and invertible. The fundamental observation is that the solution vector
\begin{equation}
    \bm x^\star = A^{-1}\bm b
\end{equation}
is identical to the vector of averages 
\begin{equation}
    x^*_i = \mu_i =\int \prod_j dx_j~ x_i p(\bm x)\ ,
\end{equation}
of the following multivariate Gaussian\footnote{For $p(\bm x)$ to be normalizable, we need $A$ to also be positive definite. In our context, the matrix $A$ is $(\mathds{1}-\alpha G)$ (see \eqref{defKs}), which is symmetric and diagonally dominant (at least on average) with positive diagonal entries, therefore it is typically invertible and positive definite by the  Gershgorin–Hadamard theorem. }
\begin{equation}
    p(\bm x)= \frac{1}{Z}\exp\left[-\frac{1}{2}\bm x^T A\bm x+\bm b^T \bm x\right]\ .\label{multivGauss}
\end{equation}

This follows from
\begin{equation}
    (\bm x-\bm x^\star)^T A(\bm x-\bm x^\star)=\bm x^T A\bm x-2\bm b^T \bm x+\bm b^T A^{-1}\bm b\ ,
\end{equation}
which allows us to write the multivariate Gaussian with mean vector $\bm x^\star$ in the form of Eq. \eqref{multivGauss}
\begin{align}
\nonumber p(\bm x) &=\frac{1}{Z'}\exp\left[-\frac{1}{2}(\bm x-\bm x^\star)^T A (\bm x-\bm x^\star)\right]\\
&=\frac{1}{Z}\exp\left[-\frac{1}{2}\bm x^T A\bm x+\bm b^T \bm x\right]\ ,
\end{align}
with $Z=Z'\exp[(1/2)\bm b^T A^{-1}\bm b]$.

Therefore
\begin{align}
    x_i^\star=\mu_i =\int dx_i~x_i p_i(x_i)\ ,\label{mui}
\end{align}
where
\begin{equation}
    p_i(x_i)=\int \prod_{j\neq i}dx_j~ p(\bm x)
\end{equation}
is the marginal distribution of the variable $x_i$ alone. Writing the solution in the form of Eq. \eqref{mui} transfers the problem from the linear algebra domain to the probabilistic domain, allowing us to tackle it with a more powerful and broader set of tools.

From now on, we further assume that the matrix $A$ of coefficients of the linear system defines a locally tree-like graph structure, where the unknowns $x_i$ live on the $N$ nodes of a graph, and the coefficients $A_{ij}\neq 0$ stand for the weight of the edge connecting node $i$ and $j$.

If the graph is a tree -- but the treatment below works very well for tree-like structures -- we can appeal to the 
GaBP scheme \cite{GaussianBP,linearshental,linsyst} -- a particular incarnation of the \emph{cavity method} \cite{cavity1,cavity2,cavity3,cavity4} from the theory of disordered systems, and of \emph{message passing} algorithms \cite{mess1,mess2,mess3} -- to find efficient and fast recursive equations for the averages $\mu_i$ we are after. Among the many virtues of the scheme is the fact that -- when the algorithm converges -- it is guaranteed to converge to the true averages (i.e. the inference is guaranteed to be exact) \cite{GaussianBP,linsyst}. In our case, the convergence of the algorithm follows from the condition \eqref{alphalambda}, which defines a \emph{walk-summable} problem (see \cite{convergenceBP}, Proposition 2).

Let us start by rewriting the marginal $p_i(x_i)$ as follows

\begin{align}
\nonumber &  p_i(x_i) =\frac{1}{Z_i}\int \prod_{j\neq i}dx_j\exp\left[-\frac{1}{2}\sum_i x_i\sum_{j\in\partial i}A_{ij}x_j+\sum_k b_k x_k\right]\\
\nonumber &=\frac{1}{Z_i}e^{-\frac{1}{2}A_{ii}x_i^2+b_i x_i}
    \int \prod_{j\in\partial i}dx_j\exp\left[- x_i\sum_{j\in\partial i}A_{ij}x_j\right]\times\\
    &\times p^{(i)}(\bm x_{\partial i})\ ,
\end{align}

where $\partial i$ denotes the set of nodes $j$ connected to $i$ ($A_{ij}\neq 0$), while $p^{(i)}(\bm x_{\partial_i})$ denotes the \emph{cavity distribution}, namely the joint distribution of remaining variables (so, from the $j$-th variable outwards) after the node $i$ has been removed from the picture.

\begin{figure}[h]
    \centering
    \fbox{\includegraphics[scale = 0.25]{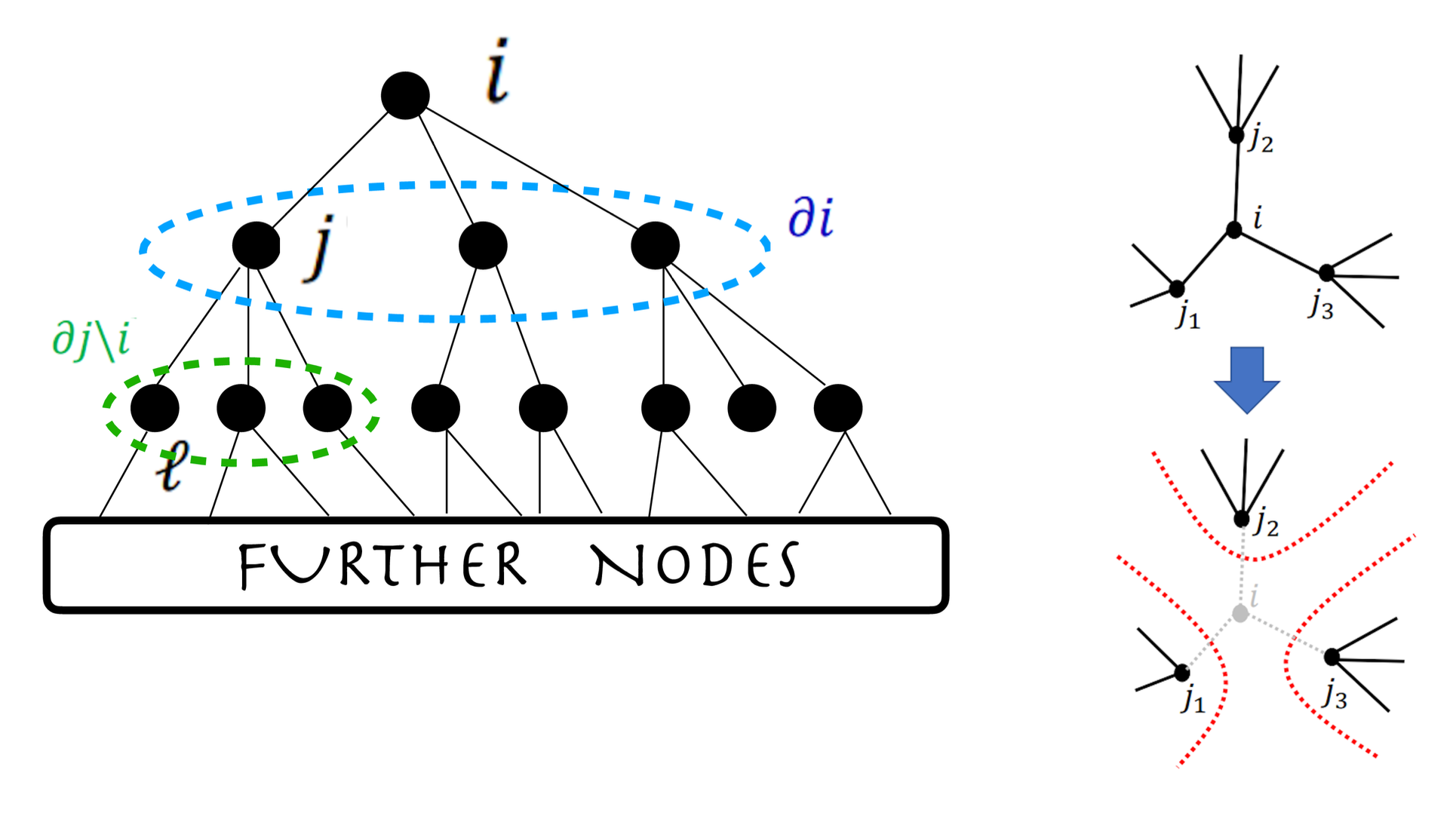}}
    \caption{Sketch of the tree structure with the node $i$ on top, the neighborhood $\partial i$ in dashed blue, and the further-down neighborhood $\partial j\setminus i$ in dashed green (left). On the right, schematic representation of the removal of node $i$, which leaves nodes $j_1$, $j_2$ and $j_3$ independent.
     }
    \label{fig:cavity}
\end{figure}

Now, in a tree structure, the nodes $j$ in the neighborhood of $i$ are only connected to each other via the node $i$ (see sketch in Fig. \ref{fig:cavity}). When the node $i$ is removed, the variables defined on these nodes become therefore independent, i.e. the cavity distribution factorizes over the nodes in the neighborhood of $i$
\begin{equation}
    p^{(i)}(\bm x_{\partial i})=\prod_{j\in \partial i}
    p_j^{(i)}(x_j)\ .
\end{equation}
Therefore
\begin{align}
  \nonumber  p_i(x_i)&=\frac{1}{Z_i}e^{-\frac{1}{2}A_{ii}x_i^2+b_i x_i}
    \times\\
    &\times\prod_{j\in \partial i}\int dx_j \exp\left[- x_i A_{ij}x_j\right]p_j^{(i)}(x_j)\ .\label{firstcavity}
\end{align}
 
We can repeat the reasoning for the cavity distribution itself
\begin{align}
  \nonumber  p_j^{(i)}(x_j)&=\frac{1}{Z_j^{(i)}}e^{-\frac{1}{2}A_{jj}x_j^2+b_j x_j}
    \times\\
    &\times\prod_{\ell\in \partial j\setminus i}\int dx_\ell\exp\left[- x_j A_{j\ell}x_\ell\right]p_\ell^{(j)}(x_\ell)\ ,\label{recursioncavity}
\end{align}
where $\partial j\setminus i$ denotes the set of neighbors of node $j$ excluding the node $i$. Note that Eq. \eqref{recursioncavity} is now a closed recursion for the cavity distributions $p_j^{(i)}$, whereas \eqref{firstcavity} is not a closed recursion for the marginal $p_i(x_i)$. Knowing the cavity marginals (solutions of \eqref{recursioncavity}), though, it is possible to compute the marginals using \eqref{firstcavity}, as we show below.

We make the (normalized) Gaussian ansatz for the cavity distribution
\begin{equation}
    p_j^{(i)}(x)=\frac{1}{Z_{j}^{(i)}} \exp\left(-\frac{(x-\mu_j^{(i)})^2}{2 V_{j}^{(i)}}\right)\label{ansatzcavityV2}
\end{equation}
with cavity mean $\mu_j^{(i)}$ and cavity variance $V_j^{(i)}$. Inserting this ansatz on the r.h.s. of \eqref{recursioncavity}, we compute the resulting Gaussian integral using the result 
\begin{equation}
    \langle e^{-M x}\rangle_{\mathcal{N}(\mu,V)}=e^{\frac{M^2 V}{2}-M \mu}\ ,
\end{equation}
where $\langle\cdot\rangle$ stands for averaging over a normalized Gaussian $\mathcal{N}(\mu,V)$ with mean $\mu$ and variance $V$. Specializing to
\begin{equation}
    M = x_j A_{j\ell}
\end{equation}
from \eqref{recursioncavity}, we see that the exponent in the r.h.s. becomes
\begin{align}
 \nonumber    &-\frac{1}{2}x_j^2\left(A_{jj}-\sum_{\ell\in\partial j\setminus i}V_\ell^{(j)}A_{j\ell}^2\right)+\\
    &+x_j \left(b_j-\sum_{\ell\in\partial j\setminus i}A_{j\ell}\mu_\ell^{(j)}\right)\ .
\end{align}

Furthermore, the average and variance of a normalized Gaussian of the form appearing in the r.h.s. of \eqref{recursioncavity}, namely
\begin{equation}
    p(x)=\frac{1}{Z}e^{-\frac{1}{2}C x^2+D x}
\end{equation}
are respectively
\begin{align}
 V &= \frac{1}{C}\\
    \mu &=\frac{D}{C}=DV\ .
\end{align}
Using the expressions above, we get -- equating mean and variance -- from \eqref{recursioncavity} and using the ansatz \eqref{ansatzcavityV2}
\begin{align}
V_j^{(i)} &=\frac{1}{A_{jj}-\sum_{\ell\in\partial j\setminus i}V_\ell^{(j)}A_{j\ell}^2}\label{sc2}\\
    \mu_j^{(i)} &=V_j^{(i)}\left(b_j-\sum_{\ell\in\partial j\setminus i}A_{j\ell}\mu_\ell^{(j)}\right)\label{sc1}\ .
\end{align}

Similarly, we make the (normalized) Gaussian ansatz for the marginal distribution
\begin{equation}
    p_j(x)=\frac{1}{Z_{j}} \exp\left(-\frac{(x-\mu_j)^2}{2 V_{j}}\right)\label{ansatzcavityV3}
\end{equation}
with mean $\mu_j$ and variance $V_j$. Inserting again the Gaussian ansatz \eqref{ansatzcavityV2} for the cavity marginal in the r.h.s. of \eqref{firstcavity}, and comparing with the ansatz \eqref{ansatzcavityV3} for the l.h.s., we obtain the following equations

\begin{align}
V_j &=\frac{1}{A_{jj}-\sum_{\ell\in\partial j}V_\ell^{(j)}A_{j\ell}^2}\label{final2}\\
    \mu_j &=V_j\left(b_j-\sum_{\ell\in\partial j}A_{j\ell}\mu_\ell^{(j)}\right)\label{final1}\ .
\end{align}
Solving the self-consistency equations \eqref{sc1} and \eqref{sc2} on the cavity graph and inserting the results into \eqref{final1} and \eqref{final2} provides the solution $x_i^\star=\mu_i$ of the linear system \eqref{linearsys}. The equations above are identical to those provided in \cite{linearshental},
after some rewriting and rearrangements. In the next section, we are going to specialize these results to the case of the linear system \eqref{defKs} defining the shifted Katz centrality on a single network instance.

\section{Katz centrality on single instance of a random graph}\label{sec:KatzSI}

To apply the formalism developed in the previous section to the Katz centrality, we may define from \eqref{defKs} the matrix $A$ as
\begin{equation}
    A_{j\ell}=\delta_{j\ell}-\alpha G_{j\ell}=
    \begin{cases}
    -\alpha &\qquad\text{if }j\neq \ell\\
    1&\qquad\text{if }j= \ell
    \end{cases}\ ,
\end{equation}
since we assume that a link exists between node $j$ and $\ell$, and that there are no self-loops. Also, $b_j =1$ for all $j$.

\begin{figure}[h]
    \centering
    \fbox{\includegraphics[scale = 0.24]{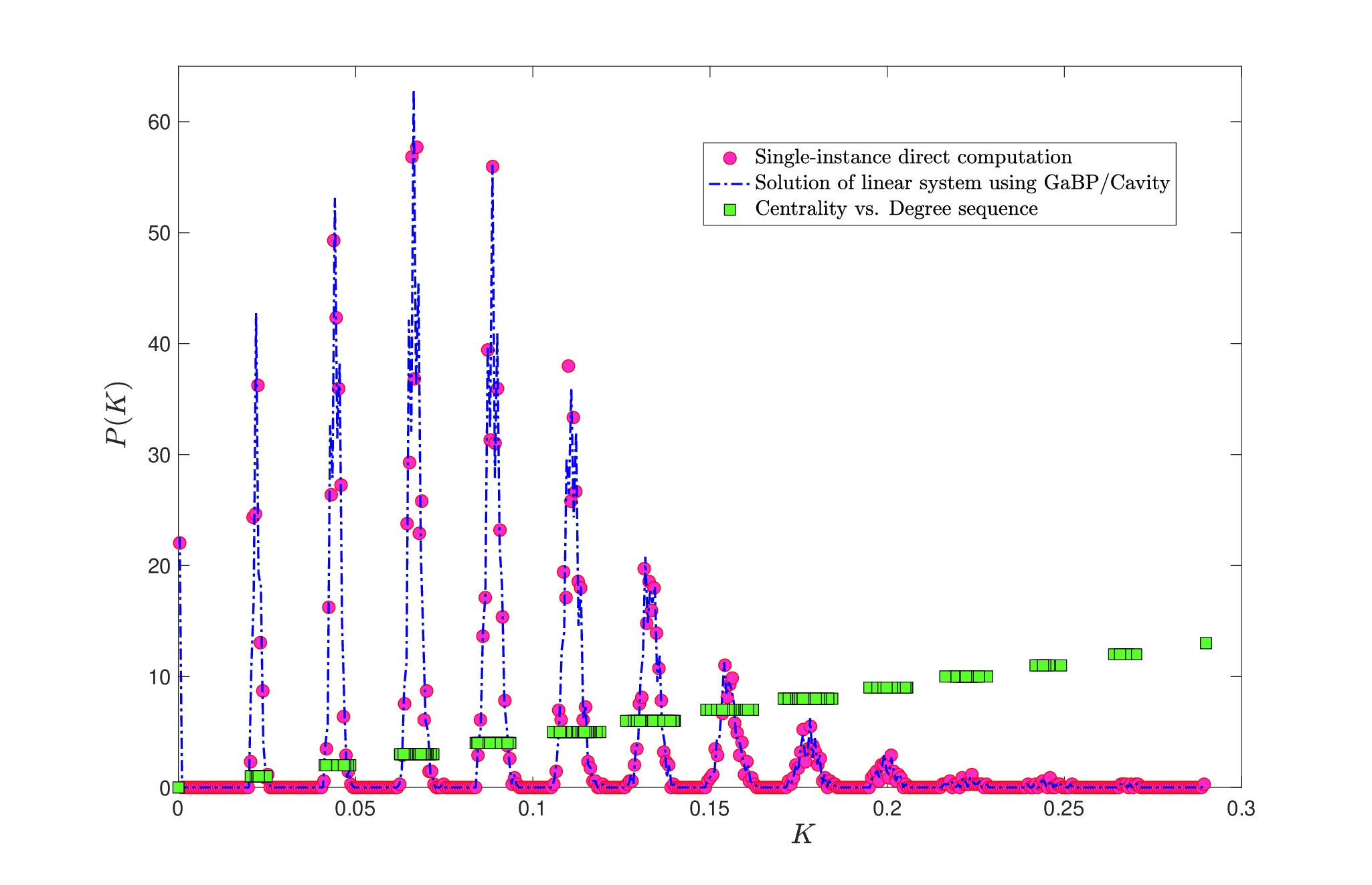}}
    \caption{Probability density function $P(K)$ of the Katz centrality with $\alpha=1/50$ computed over a single instance of an Erd\H os-R\'enyi graph of size $N=5000$ with average degree $c=4$ by direct matrix inversion from Eq. \eqref{defKatzvector} (pink circles). Blue dot-dashed line: 
    GaBP/cavity solution of the linear system as given in Eqs. \eqref{KatzVariance1}, \eqref{sc1Katz}, \eqref{Katzfinal2}, \eqref{Katzfinal1} and \eqref{Kofmu}. The coordinates $(K_j,k_j)$ of each green square $j=1,\ldots,N$ provide the degree $k_j$ of node $j$ against its centrality $K_j$. 
     }
    \label{fig:single}
\end{figure}

\begin{figure}[h]
    \centering
    \fbox{\includegraphics[scale = 0.24]{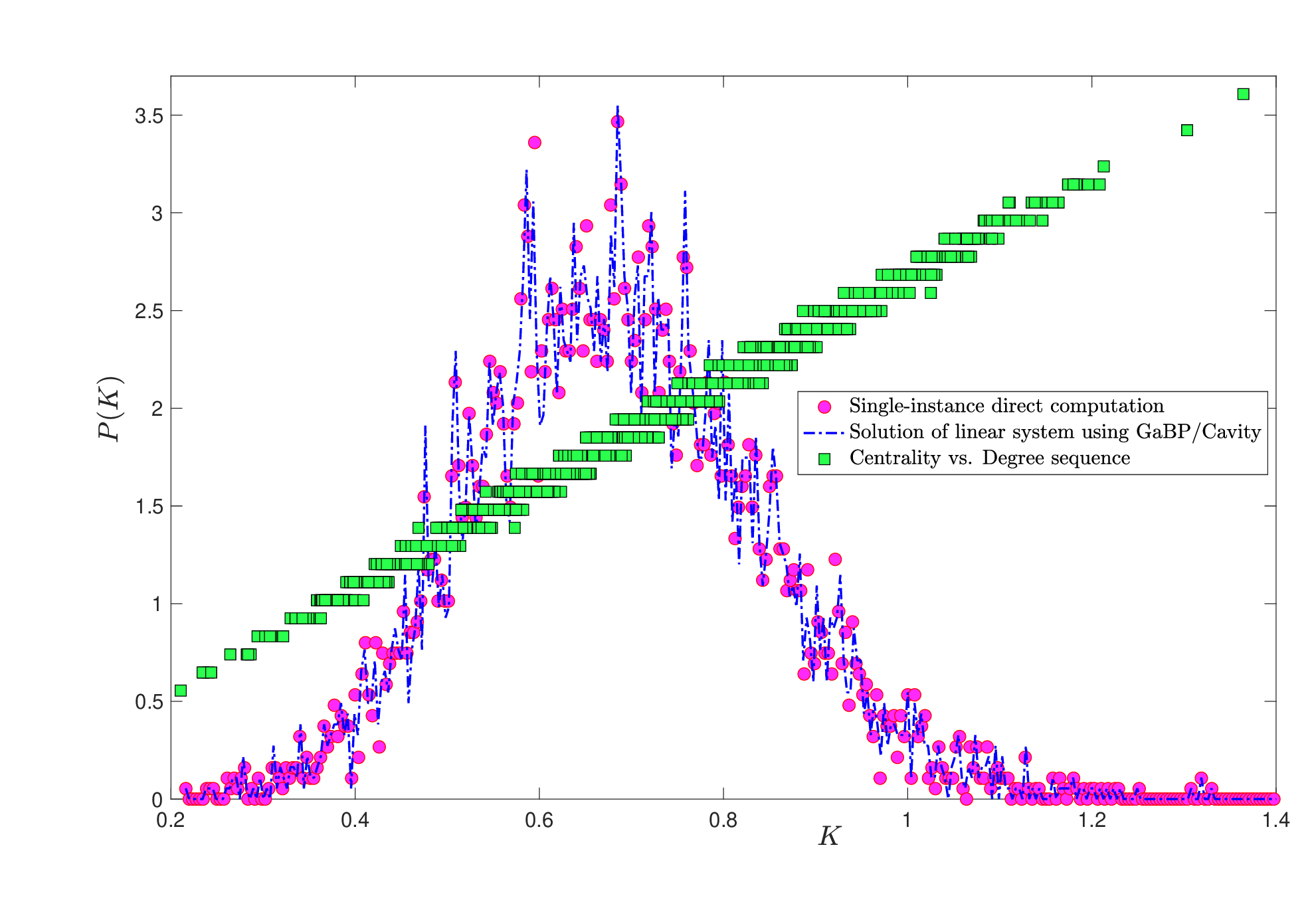}}
    \caption{Probability density function $P(K)$ of the Katz centrality with $\alpha=1/50$ computed over a single instance of an Erd\H os-R\'enyi graph of size $N=5000$ with average degree $c=20$ by direct matrix inversion from Eq. \eqref{defKatzvector} (pink circles). Blue dot-dashed line: 
    GaBP/cavity solution of the linear system as given in Eqs. \eqref{KatzVariance1}, \eqref{sc1Katz}, \eqref{Katzfinal2}, \eqref{Katzfinal1} and \eqref{Kofmu}. The coordinates $(K_j,k_j/m)$ of each green square $j=1,\ldots,N$ provide the degree $k_j$ of node $j$ -- rescaled by a factor $m=10.81$ to make it visible on the same scale -- against its centrality $K_j$. 
     }
    \label{fig:single2}
\end{figure}

The self-consistent cavity equations thus become

\begin{align}
 V_j^{(i)} &=\frac{1}{1-\alpha^2\sum_{\ell\in\partial j\setminus i}V_\ell^{(j)}}\label{KatzVariance1}\\
\mu_j^{(i)} &=V_j^{(i)}\left(1+\alpha\sum_{\ell\in\partial j\setminus i}\mu_\ell^{(j)}\right)\label{sc1Katz}\\ 
V_j &=\frac{1}{1-\alpha^2\sum_{\ell\in\partial j}V_\ell^{(j)}}\label{Katzfinal2}\\
    \mu_j &=V_j\left(1+\alpha\sum_{\ell\in\partial j}\mu_\ell^{(j)}\right)\label{Katzfinal1}\ ,
\end{align}
from which the Katz centrality $K_i$ of node $i$ can be efficiently determined from \eqref{defKs} as
\begin{equation}
    K_i = \mu_i - 1\ .\label{Kofmu}
\end{equation}

In Fig. \ref{fig:single}, we plot the Katz centrality distribution for a single instance of an Erd\H os-R\'enyi graph of size $N=5000$ with average degree $c=4$, along with the GaBP/cavity solution of the recursions above, as well as the degree sequence staircase (green squares). From the plot, one easily infers that the centrality distribution is naturally decomposed into contributions (peaks) coming from nodes of different degrees. Increasing the average connectivity $c$, the peaks would gradually merge, as more and more nodes of different degrees happen to have the same centrality (see Fig. \ref{fig:single2} for $c=20$).

As a simple further check of the formalism, we may specialize these equations to the case of a random regular graph having all nodes with the same degree, $p(k)=\delta_{k,c}$. The Katz centrality of all nodes is the same, and given by
\begin{equation}
    K_i = \frac{1}{1-\alpha c}-1\qquad\forall i\ ,\label{KatzFinalRRG}
\end{equation}
(see Lemma 3.1 in \cite{paton}).

The set of recursive equations above specializes to
\begin{align}
 V &=\frac{1}{1-\alpha^2(c-1)V}\label{KatzVariance1RRG}\\
\mu &=V\left(1+\alpha(c-1)\mu\right)\label{sc1KatzRRG}\\ 
\tilde V &=\frac{1}{1-\alpha^2 c V}\label{Katzfinal2RRG}\\
    \tilde\mu &=\tilde V\left(1+\alpha c\mu\right)\label{Katzfinal1RRG1}\ ,
\end{align}
where we imposed that all cavity fields take up a single value ($\mu$ and $V$) on every edge, and similarly for the marginal fields ($\tilde\mu$ and $\tilde V$). The equations above can be easily solved, and the value of $\tilde\mu = 1/(1-\alpha c)$. It follows therefore from \eqref{Kofmu} that the Katz centrality of nodes in a random regular graph indeed comes out as \eqref{KatzFinalRRG}.

\section{Probability $P(K)$ over the ensemble}\label{sec:PK}

We are now interested in leveraging the results of the previous section -- valid for a single instance of a random network -- to compute the probability density function $P(K)$ of finding a node $i$ with centrality $P(K)=\mathrm{Prob}[K_i=K]$ in an ensemble of large undirected random graphs. Going from single-instance cavity results to distributions over an ensemble is a quite standard procedure (see \cite{Vivoreview} for a review), which we report here for completeness.

First, one has to focus on the joint probability density function $\pi(\mu,V)$ of observing a cavity mean $\mu_j^{(i)}=\mu$ and a cavity variance $V_j^{(i)}=V$ in the ensemble. To do so, one observes that the self-consistency equations for the cavity variance and mean (\eqref{KatzVariance1} and \eqref{sc1Katz}) refer to the \emph{links} of the underlying graph. In an infinitely large network, links can be distinguished from one another by the degree of the node they are pointing to. Therefore, considering an edge $(i,j)$ pointing to a node $j$ of degree $k$, the value $(\mu,V)$ of the pair formed by the cavity mean $\mu_{j}^{(i)}$ and the cavity variance $V_j^{(i)}$ -- both living on this edge -- is determined by the set $\left\{\mu_{\ell},V_\ell\right\}_{k-1}$ of the $k-1$ values of the pair $(\mu_{\ell}^{(j)},V_{\ell}^{(j)})$ living on each of the edges connecting $j$ with its neighbors $\ell\in\partial j \backslash i$. In an infinite system, these values can be regarded as $k-1$ independent realizations of the pair of random variables of type $\mu_{\ell}^{(j)}$ and $V_{\ell}^{(j)}$, each drawn from the same joint pdf $\pi(\mu,V)$.

The joint pdf $\pi(\mu,V)$ is then obtained by averaging the contributions coming from every link w.r.t. the probability $\frac{k}{c}p(k)$ of having a link pointing to a node of degree $k$\footnote{It can be observed that in general the probability that a node of degree $k$ is connected to a node of degree $k^\prime$ is conditional, namely $P(k^\prime|k)$. However, configuration model ensembles (including the Erd\H{o}s-R\'enyi ensemble) are cases of random uncorrelated networks, hence $P(k^\prime|k)$ is independent of $k$. Therefore, $P(k^\prime|k)$ reduces to the probability that an edge points to a node of degree $k^\prime$, which can be defined as the ratio between the number of edges pointing to nodes of degree $k'$ , $k^\prime p(k^\prime)$, and the number of edges pointing to nodes of any degree, i.e. the sum $\sum_{k^\prime}k^\prime p(k^\prime)=c$.}, with $p(k)$ being the degree distribution of the network, and $c\sim\mathcal{O}(1)$ the average connectivity. This reasoning leads to the self-consistency equation
\begin{widetext}
\begin{equation}
\pi(\mu,V)=\sum_{k=1}^{\infty}p(k)\frac{k}{c}\int \{d\pi\}_{k-1} \delta\left(\mu-V\left(1+\alpha\sum_{\ell=1}^{k-1}\mu_\ell\right)\right)\delta\left(V-\frac{1}{1-\alpha^2\sum_{\ell=1}^{k-1}V_\ell}\right)\ ,
\label{eq:ch1_cavity_pi}
\end{equation}
\end{widetext}
where  $\{d\pi\}_{k-1}=\prod_{\ell=1}^{k-1} d\mu_\ell dV_\ell \pi(\mu_\ell,V_\ell)$. The recursive distributional equation \eqref{eq:ch1_cavity_pi} can be efficiently solved via a population dynamics algorithm (see Section \ref{sec:popdyn}). Note that the integral equations above can now be considered and solved independently of the network problem that originated them, since no other information about the topology of such network enters the picture apart from the degree distribution $p(k)$, which makes this approach so general and powerful.

The same reasoning can be applied to find the joint pdf $\tilde\pi(\tilde\mu,\tilde V)$ of the pair $(\mu_i,V_i)$ satisfying equations \eqref{Katzfinal2} and \eqref{Katzfinal1}. From there, one notices that the $\mu_i$ and $V_i$ are variables related to \emph{nodes}, rather than edges. Since in the infinite size limit the nodes can be distinguished from one another by their degree, the joint pdf $\tilde\pi(\tilde\mu,\tilde V)$ can be written in terms of the solution $\pi(\mu, V)$ of \eqref{eq:ch1_cavity_pi} as
\begin{widetext}
\begin{equation}
\tilde\pi(\tilde\mu,\tilde V)=\sum_{k=0}^{\infty}p(k)\int \{d\pi\}_{k}  \delta\left(\tilde\mu-\tilde V\left(1+\alpha\sum_{\ell=1}^{k}\mu_\ell\right)\right)\delta\left(\tilde V-\frac{1}{1-\alpha^2\sum_{\ell=1}^{k}V_\ell}\right) \ ,
\label{eq:pitilde}
\end{equation}
\end{widetext}
where $p(k)$ is again the degree distribution. Note that the r.h.s. of \eqref{eq:pitilde} is a sum of $k$-fold integrals involving $\pi$ and not $\tilde\pi$, because $\mu_i$ and $V_i$ are defined in terms of the ``cavity'' pair (see Eqs. \eqref{Katzfinal2} and \eqref{Katzfinal1}). Also, the integral relations above evidently preserve the normalization of the joint pdfs $\pi$ and $\tilde\pi$.

After solving \eqref{eq:pitilde} for the joint pdf $\tilde\pi(\tilde\mu,\tilde V)$ of the variables of type $\mu_i$ and $V_i$, we appeal to Eq. \eqref{Kofmu} and the definition of the shifted Katz centrality as a linear system in Eq. \eqref{defKs} to write the pdf $P(K_s)$ as
\begin{equation}
    P(K_s) = \int d\tilde V~ \tilde\pi(K_s,\tilde V)\ ,
\end{equation}
from which we readily get
\begin{equation}
P(K) = \int d\tilde V~ \tilde\pi(K+1,\tilde V)=\sum_{k=0}^\infty p(k)P(K|k)\ ,\label{PKsuperposition}
\end{equation}
with the pdf $P(K|k)$ of a node having centrality $K$ given that it has degree $k$ given by
\begin{widetext}
\begin{equation}
P(K|k)=\int \{d\pi\}_{k}  \delta\left(K+1- \left(\frac{1}{1-\alpha^2\sum_{\ell=1}^{k}V_\ell}\right)\left(1+\alpha\sum_{\ell=1}^{k}\mu_\ell\right)\right)\ .
\end{equation}
\end{widetext}
Written as in Eq. \eqref{PKsuperposition}, the pdf of the Katz centrality is naturally expressed as a superposition of contributions, each coming from nodes of degree $k$. For sufficiently low average connectivity $c$, the individual degree-$k$ contributions are clearly visible in the form of distinct peaks (see e.g. Fig. \ref{fig:c4} and \ref{fig:c10} below).

\section{Numerical solution using population dynamics}\label{sec:popdyn}
In this section, we describe the stochastic population dynamics algorithm that leads to the solution of the self-consistency equation \eqref{eq:ch1_cavity_pi} for the joint pdf $\pi(\mu,V)$, coupled with the sampling procedure to evaluate \eqref{eq:pitilde}. This kind of algorithm is widely used in the study of amorphous systems \cite{zippelius}, spin glasses \cite{PopDyn1,PopDyn2},  random matrices \cite{Vivoreview,kuehn,vivo1,vivo2} and percolation in sparse networks \cite{rogersperc}.

First, in order to solve \eqref{eq:ch1_cavity_pi}, one represents the joint pdf $\pi(\mu,V)$ in terms of two populations of $N_P$ real values, ${\bf M} \equiv\{\mu_i\}$ and ${\bf V}\equiv\{V_i\geq 0\}$ for $i=1,\ldots,N_P$, which are assumed to be sampled from that joint pdf. Given that the true jpdf is initially unknown, a starting population is initialized randomly. 

Similarly, one represents the joint pdf $\tilde\pi(\tilde\mu,\tilde V)$ in terms of two populations of $N_P$ real values, ${\bf \tilde M} \equiv\{\tilde\mu_i\}$ and ${\bf \tilde V}\equiv\{\tilde V_i\geq 0\}$ for $i=1,\ldots,N_P$, which are assumed to be sampled from that joint pdf. Again, a starting population is initialized randomly. 

Then the following stochastic algorithm is iterated until two stable populations are reached:
\begin{enumerate}
\item Generate a random integer $k$ from the distribution $\frac{k}{c}p(k)$, where $p(k)$ is the degree distribution of the ensemble of interest and $c=\sum_k k p(k)$ is the average degree;
\item Generate a random integer $\tilde k$ from the degree distribution $p(k)$;
\item Select $k-1$ elements $\mu_\ell^{(old)}$ at random from the population $\bf M$, and $k-1$ elements $V_\ell^{(old)}$ from the population $\bf V$;
\item Select $\tilde k$ elements $\tilde\mu_\ell^{(old)}$ at random from the population $\bf \tilde M$, and $\tilde k$ elements $\tilde V_\ell^{(old)}$ from the population $\bf \tilde V$;
\item Compute the new values
\begin{align}
V^{(new)} &=\frac{1}{1-\alpha^2\sum_{\ell=1}^{k-1}V_\ell^{(old)}}\\
\mu^{(new)} &=V^{(new)}\left(1+\alpha\sum_{\ell=1}^{k-1}\mu_\ell^{(old)}\right)\\
\tilde V^{(new)} &=\frac{1}{1-\alpha^2\sum_{\ell=1}^{\tilde k}V_\ell^{(old)}}\\
\tilde\mu^{(new)} &=\tilde V^{(new)}\left(1+\alpha\sum_{\ell=1}^{\tilde k}\mu_\ell^{(old)}\right)\ .
\end{align}
\item Replace a randomly selected element $V^{(old)}$ of $\bf V$ with $V^{(new)}$, and a randomly selected element $\mu^{(old)}$ of $\bf M$ with $\mu^{(new)}$.
\item Replace a randomly selected element $\tilde V^{(old)}$ of $\bf \tilde V$ with $\tilde V^{(new)}$, and a randomly selected element $\tilde \mu^{(old)}$ of $\bf \tilde M$ with $\tilde\mu^{(new)}$.
\item Return to 1. 
\end{enumerate}
Once two stable populations are reached, the pdf of the shifted centrality is simply obtained by histogramming the population $\bf\tilde M$. The fact that the populations have reached convergence is established by monitoring the first and second moments of the samples and stopping when they have clearly plateaued. 

In the following, we show the comparison between the numerical solution obtained with population dynamics and direct matrix inversion for Erd\H{o}s-Renyi and scale-free networks. Erd\H{o}s-Renyi networks were built by drawing each possible link with the same probability $p=c/(N-1)$, which leads to networks with a Poisson degree distribution in the limit of large $N$. Scale-free networks were built using the uncorrelated configuration model \cite{generationSF}: Each node was assigned a number of half-links drawn from a power law distribution $P(k)\propto k^{-\gamma}$, and these were randomly matched to form links. With this procedure, we avoided the occurrence of multiple links and self-loops. Furthermore, to prevent degree correlations we imposed a cut-off to the degree sequence so that the maximum allowed degree is $\sqrt{k_{min} N}$, with $k_{min}$ being the minimum degree.

To produce the figures below, we use the following parameters: 
\begin{itemize}
\item for E-R networks (Fig. \ref{fig:c4}, \ref{fig:c10}, \ref{fig:c35}) $N_P=10^5$ for the population dynamics, and $100$ sweeps (meaning that each population member has been updated $100$ times on average), with $\alpha=1/40$ and different values $c=4,10,35$ for the average connectivity. We also perform direct matrix inversion on the adjacency matrices of $1000$ E-R networks of size $N=1000$ for $c=4,10$, while for $c=35$ we averaged over $100$ networks of size $N=10000$.
\item for Scale-Free networks (Fig. \ref{fig:gamma25}, \ref{fig:gamma3}, \ref{fig:gamma4}) $N_P=10^6$ for the population dynamics, and $100$ sweeps, with $\alpha=1/40$. The network parameters are $\gamma=2.5,3,4$ respectively, with minimal degree $k_{min}=3$ and degree cutoff at $\sqrt{N k_{min}}$ to ensure no correlation between degrees \cite{generationSF}. We perform direct matrix inversion on the adjacency matrices of $100$ Scale Free networks of size $N=10000$. 
\end{itemize}

\begin{figure}[h]
    \centering
    \fbox{\includegraphics[scale = 0.4]{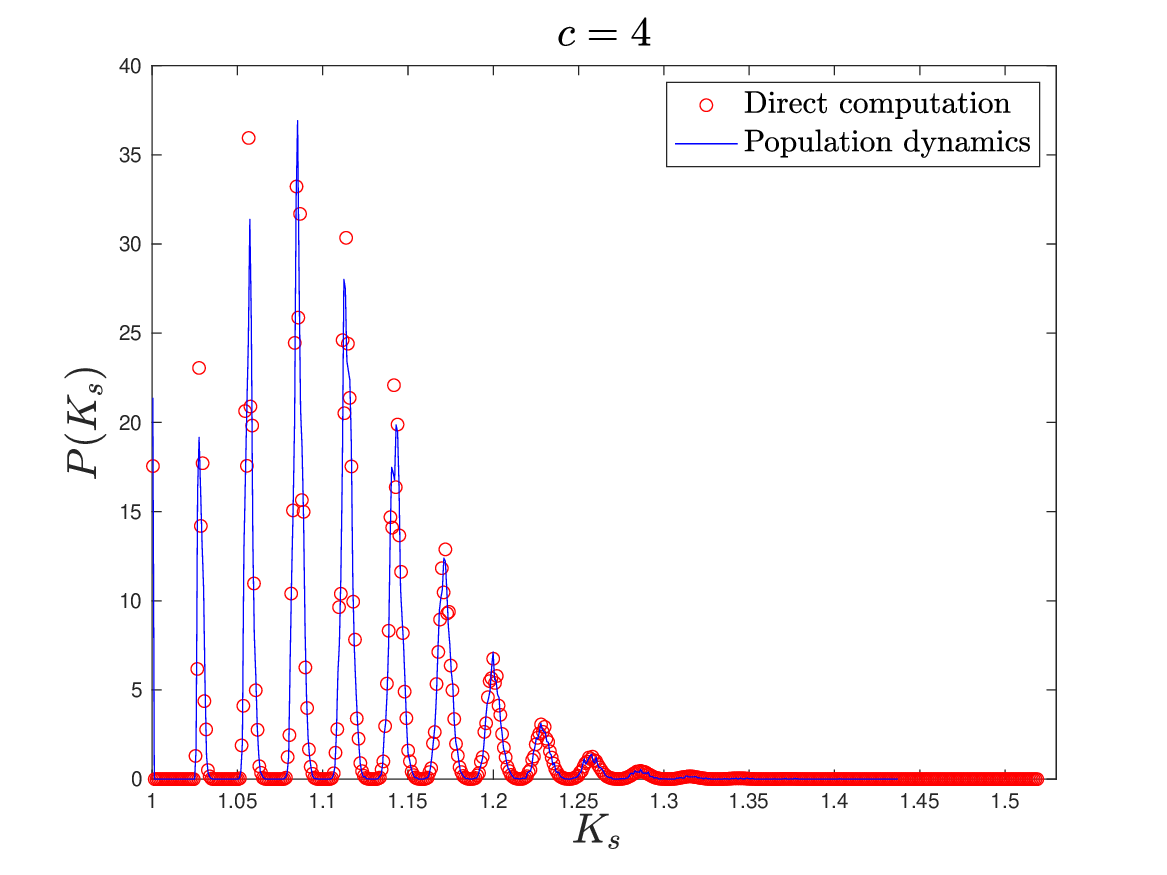}}
    \caption{Probability density function $P(K_s)$ of the shifted Katz centrality with $\alpha=1/40$ computed over an ensemble of $1000$ Erd\H os-R\'enyi graphs of size $N=1000$ with average degree $c=4$ by direct matrix inversion from Eq. \eqref{defKatzvector} (red circles). Blue solid line: distribution of the population $\bf \tilde M$ after reaching equilibrium, with $N_P=10^5$ population members and $100$ updating sweeps (see Section \ref{sec:popdyn} for details). }
    \label{fig:c4}
\end{figure}

\begin{figure}[h]
    \centering
    \fbox{\includegraphics[scale = 0.4]{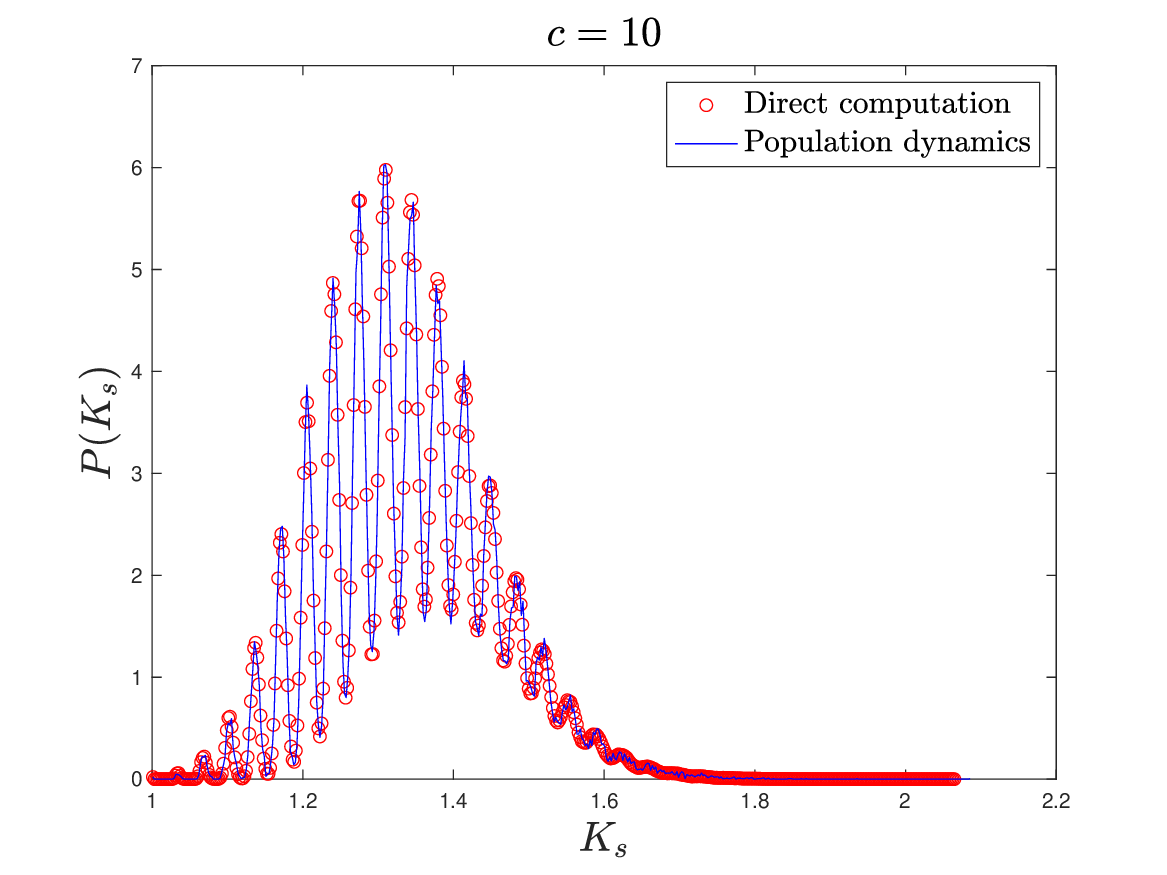}}
    \caption{Probability density function $P(K_s)$ of the shifted Katz centrality with $\alpha=1/40$ computed over an ensemble of $1000$ Erd\H os-R\'enyi graphs of size $N=1000$ with average degree $c=10$ by direct matrix inversion from Eq. \eqref{defKatzvector} (red circles). Blue solid line: distribution of the population $\bf \tilde M$ after reaching equilibrium, with $N_P=10^5$ population members and $100$ updating sweeps (see Section \ref{sec:popdyn} for details).}
    \label{fig:c10}
\end{figure}

\begin{figure}[h]
    \centering
    \fbox{\includegraphics[scale = 0.4]{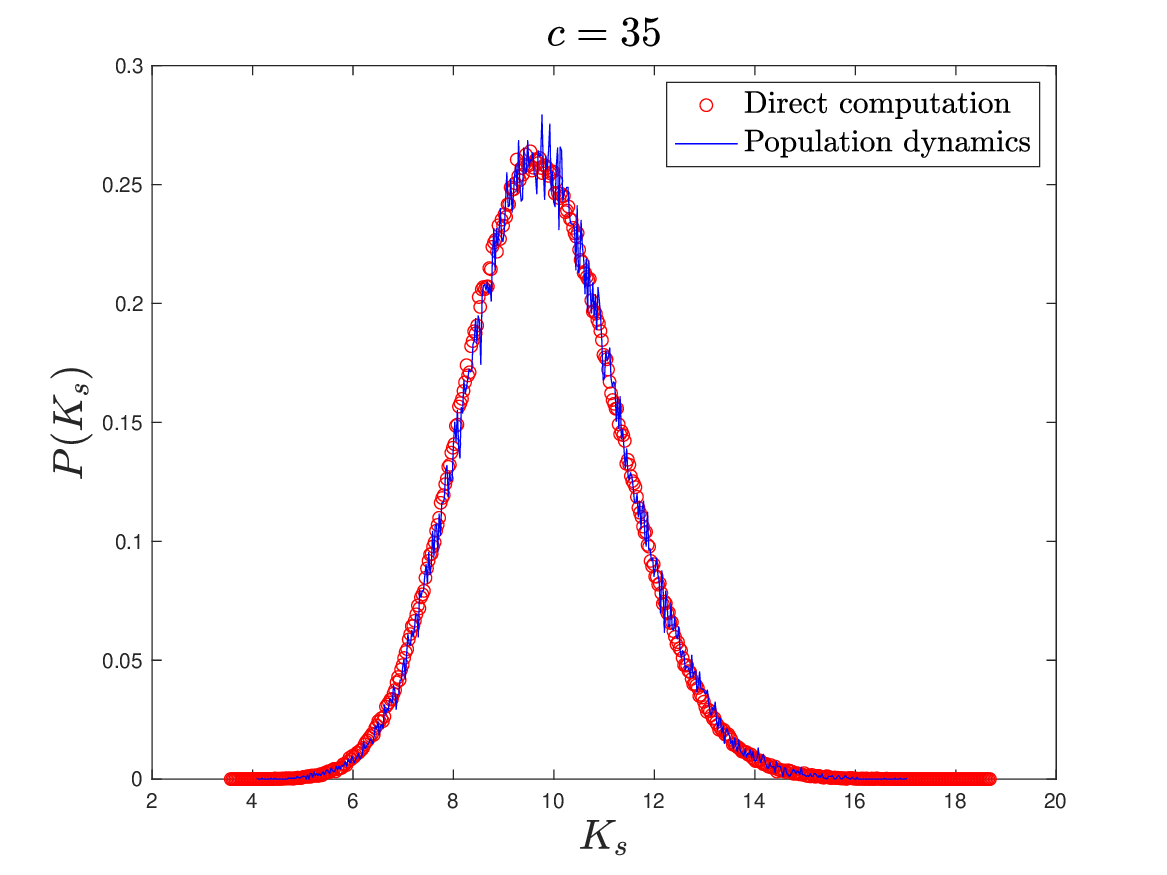}}
    \caption{Probability density function $P(K_s)$ of the shifted Katz centrality with $\alpha=1/40$ computed over an ensemble of $100$ Erd\H os-R\'enyi graphs of size $N=10000$ with average degree $c=35$ by direct matrix inversion from Eq. \eqref{defKatzvector} (red circles). Blue solid line: distribution of the population $\bf \tilde M$ after reaching equilibrium, with $N_P=10^5$ population members and $100$ updating sweeps (see Section \ref{sec:popdyn} for details).}
    \label{fig:c35}
\end{figure}


\begin{figure}[h]
    \centering
    \fbox{\includegraphics[scale = 0.4]{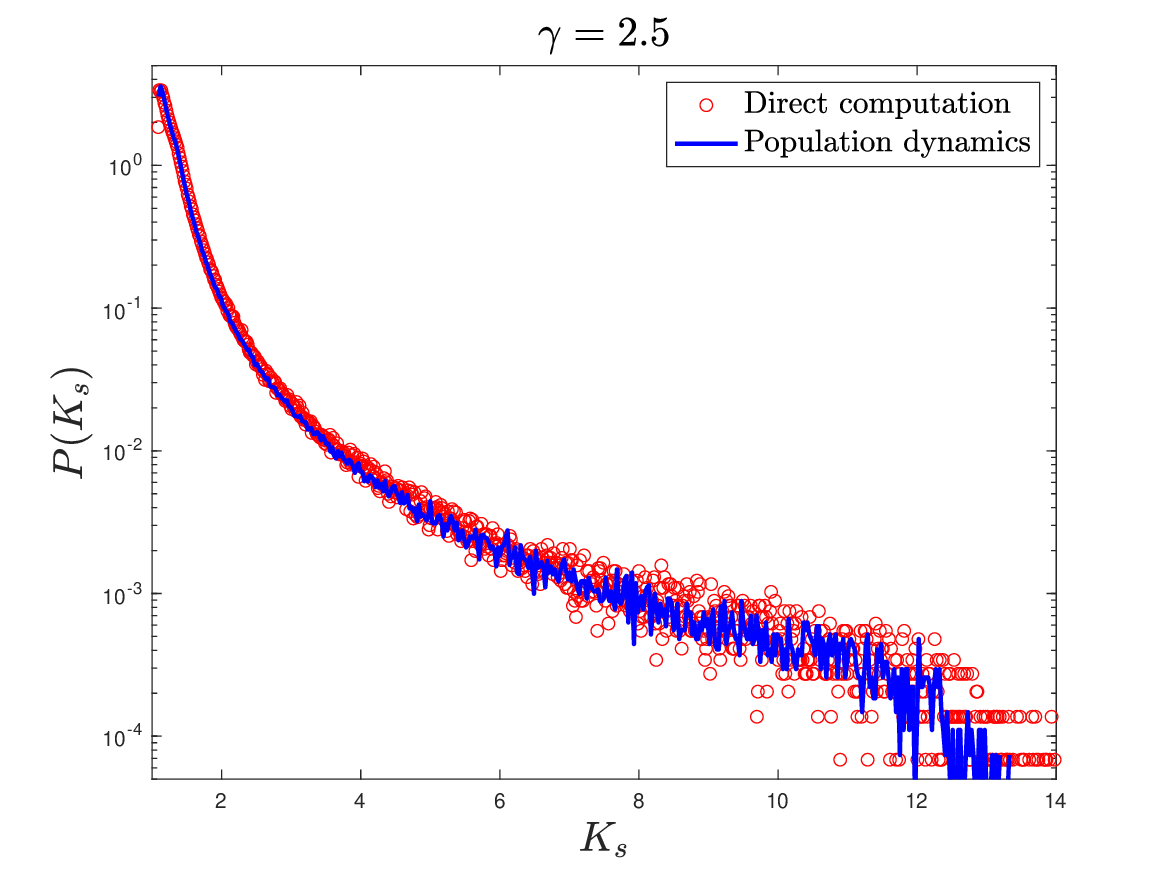}}
    \caption{Probability density function $P(K_s)$ in semi-logarithmic scale of the shifted Katz centrality with $\alpha=1/40$ computed over an ensemble of $100$ Scale Free graphs of size $N=10000$ with parameter $\gamma=2.5$ and minimum degree $k_{min}=3$ by direct matrix inversion from Eq. \eqref{defKatzvector} (red circles). Blue solid line: distribution of the population $\bf \tilde M$ after reaching equilibrium, with $N_P=10^6$ population members and $100$ updating sweeps (see Section \ref{sec:popdyn} for details).}
    \label{fig:gamma25}
\end{figure}

\begin{figure}[h]
    \centering
    \fbox{\includegraphics[scale = 0.4]{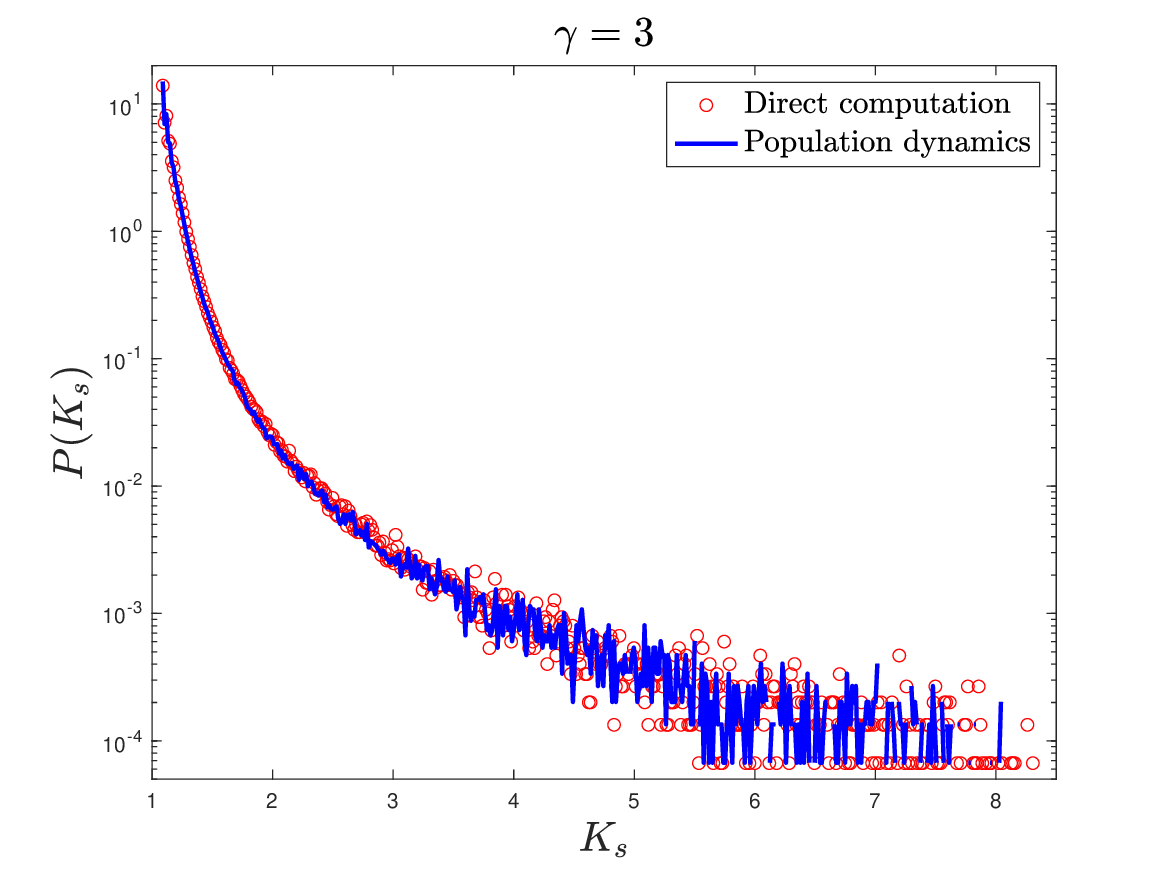}}
    \caption{Probability density function $P(K_s)$ in semi-logarithmic scale of the shifted Katz centrality with $\alpha=1/40$ computed over an ensemble of $100$ Scale Free graphs of size $N=10000$ with parameter $\gamma=3$ and minimum degree $k_{min}=3$ by direct matrix inversion from Eq. \eqref{defKatzvector} (red circles). Blue solid line: distribution of the population $\bf \tilde M$ after reaching equilibrium, with $N_P=10^6$ population members and $100$ updating sweeps (see Section \ref{sec:popdyn} for details).}
    \label{fig:gamma3}
\end{figure}

\begin{figure}[h]
    \centering
    \fbox{\includegraphics[scale = 0.4]{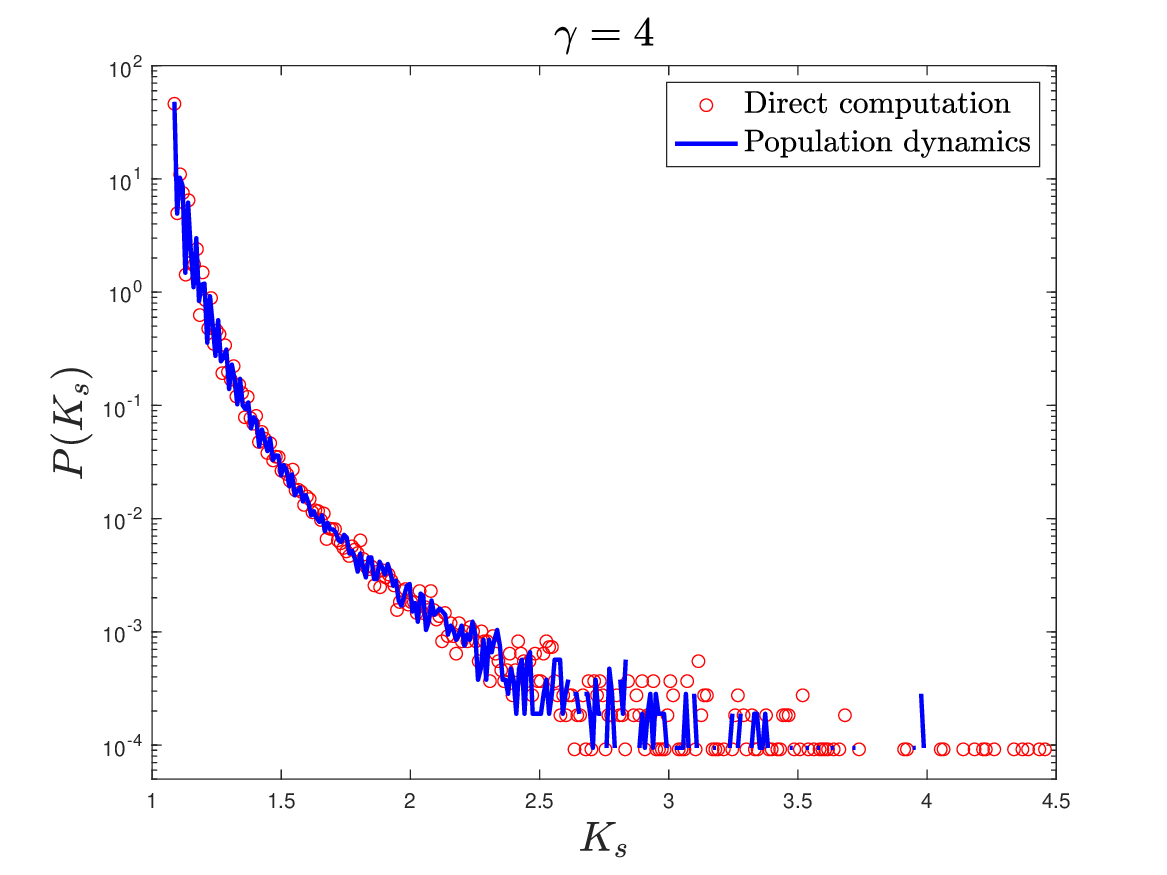}}
    \caption{Probability density function $P(K_s)$ in semi-logarithmic scale of the shifted Katz centrality with $\alpha=1/40$ computed over an ensemble of $100$ Scale Free graphs of size $N=10000$ with parameter $\gamma=4$ and minimum degree $k_{min}=3$ by direct matrix inversion from Eq. \eqref{defKatzvector} (red circles). Blue solid line: distribution of the population $\bf \tilde M$ after reaching equilibrium, with $N_P=10^6$ population members and $100$ updating sweeps (see Section \ref{sec:popdyn} for details).}
    \label{fig:gamma4}
\end{figure}

\section{Centrality distribution from rank-1 approximation}\label{sec:rank1}

In this section, we consider the rank-$1$ approximation to ranking measures proposed in \cite{bart1}, and we show that it leads to an approximate but explicit formula for the distribution $P(K_s)$, which works very well for $c$ sufficiently high.

The idea is to replace the symmetric adjacency matrix $G$ featuring in Eq. \eqref{defKatzvector} with a rank-$1$ approximation $\hat G$ defined as
\begin{equation}
    \hat G = \frac{1}{\bar k N} \bm k\bm k^T\ , 
\end{equation}
where $\bm k =\{k_1,\ldots,k_N\}^T$ is the degree sequence of the network represented by $G$, arranged in a column vector, and $\bar k$ is the mean degree $\frac{1}{N} \sum_i k_i$. Constructed in this way, the matrix $\hat G$ is rank-$1$ and has the same degree sequence (row sums) of the original matrix $G$. From Eq. \eqref{defKatzvector}, replacing $G$ with $\hat G$ and using Sherman-Morrison \cite{sherman1950} to compute the inverse matrix, we obtain
\begin{align}
\nonumber    \bm K_s\simeq (\mathds{1}-\alpha \hat G)^{-1}\bm 1 &= \left(\mathds{1}+\frac{\alpha\hat G}{1-\alpha\frac{\sum_i k_i^2}{\sum_i k_i}}\right)\bm 1\\
    &=\bm 1+\frac{\alpha}{1-\alpha \frac{\sum_i k_i^2}{\sum_i k_i }} \bm k\ .\label{eq:KsVsk}
\end{align}
Note that this rank-$1$ approximation gives a different -- and superior, as we argue below -- result from a simple linear truncation of the resolvent matrix, which would yield instead
\begin{equation}
   \bm K_s\simeq (\mathds{1}+\alpha G+\mathcal{O}(\alpha^2))\bm 1  =\bm 1+\alpha \bm k\ .\label{eq:lineartruncresolvent}
\end{equation}

To make further analytical progress, we appeal to the Law of Large Numbers for large $N$ to further approximate
\begin{align}
    \sum_i k_i &\approx N \sum_{k=0}^\infty k p(k)\equiv N c\\
    \sum_i k_i^2 &\approx N \sum_{k=0}^\infty k^2 p(k)\equiv N\overline{k^2}\ .
\end{align}
The relation \eqref{eq:KsVsk} allows us to write an approximate formula for the pdf of the Katz centrality for a large network with degree distribution $p(k)$ as
\begin{equation}\label{rank1Katz}
    P(K)\simeq \sum_{k=0}^\infty p(k)\delta\left(K-\frac{\alpha}{1-\alpha \frac{\overline{k^2}}{c}}k\right)\ .
\end{equation}

Specializing for instance to a large Erd\H os-R\'enyi network with finite mean degree\footnote{On scale-free networks with exponent $\gamma$, the second moment diverges with $N$. If we consider the structural cutoff $k_{max}\sim N^{1/2}$, we have that $\langle k^2\rangle\sim N^{(3-\gamma)/2}$. This implies that $\alpha$ should go to zero as $N$ increases for equation \eqref{rank1Katz} to be meaningful. A similar conclusion can be reached from condition \eqref{alphalambda} using the results for the maximum eigenvalue of networks generated with the configuration model reported in \cite{dionigi2023largest}.} $c$ -- characterized by a Poisson degree distribution $p(k)=e^{-c}c^k/k!$ -- we see that the centrality distribution is approximated by a Poisson-weighted Dirac comb
\begin{equation}
    P(K)\simeq \sum_{k=0}^\infty e^{-c}\frac{c^k}{k!}\delta\left(K-\frac{\alpha}{1-\alpha (1+c)}k\right)\ ,\label{eq:Diraccomb}
\end{equation}
where we used
\begin{align}
    \sum_{k=0}^\infty k \frac{e^{-c}c^k}{k!} &=c\\
    \sum_{k=0}^\infty k^2 \frac{e^{-c}c^k}{k!} &=c+c^2\ .
\end{align}
See Fig. \ref{fig:Rank1_1} and \ref{fig:Rank1_2} for a comparison between the pdf $P(K_s)$ of the shifted Katz centrality $K_s$ obtained by randomly generated Erd\H os-R\'enyi networks using the inversion formula \eqref{defKatzvector}, and the Dirac comb approximate formula \eqref{eq:Diraccomb} with $K=K_s-1$. For the simulations, we use an ensemble of $30$ Erd\H os-R\'enyi networks of size $N=5000$ with $c=30$ and $\alpha=1/45$ (Fig. \ref{fig:Rank1_1}), and $c=4$ and $\alpha=1/30$ (Fig. \ref{fig:Rank1_2}). We observe that the approximate formula \eqref{eq:Diraccomb} works very well for higher $c$ throughout the full set of allowed values of $\alpha$ (see \eqref{alphalambda}), whereas for lower $c$ -- where the actual distribution has a pronounced multi-modality -- it correctly reproduces the typical values of the centrality possessed by nodes of degree $k$ (i.e. the location of the $k$-th peak) and the value of the probability mass under each peak (magnified by a factor $20$ in Fig. \ref{fig:Rank1_2} to make the two distributions visible on the same scale). The ``network'' effect in a sparse regime therefore essentially amounts to dressing 
the degree-only information with some noise, with these fluctuations giving rise to the peaks of the centrality distribution visible in Fig. \ref{fig:Rank1_2}. Moreover, in Fig. \ref{fig:Rank1_2} we also provide the approximate Dirac comb formula that would result from using a simple linear truncation of the resolvent matrix (see \eqref{eq:lineartruncresolvent}) instead of the more sophisticated rank-$1$ approximation. We find that the simple linear truncation does not capture the location of the peaks nearly as accurately as the rank-$1$ approximation, with a clear shift of all values to the left.

\begin{figure}[h]
    \centering
    \fbox{\includegraphics[scale = 0.29]{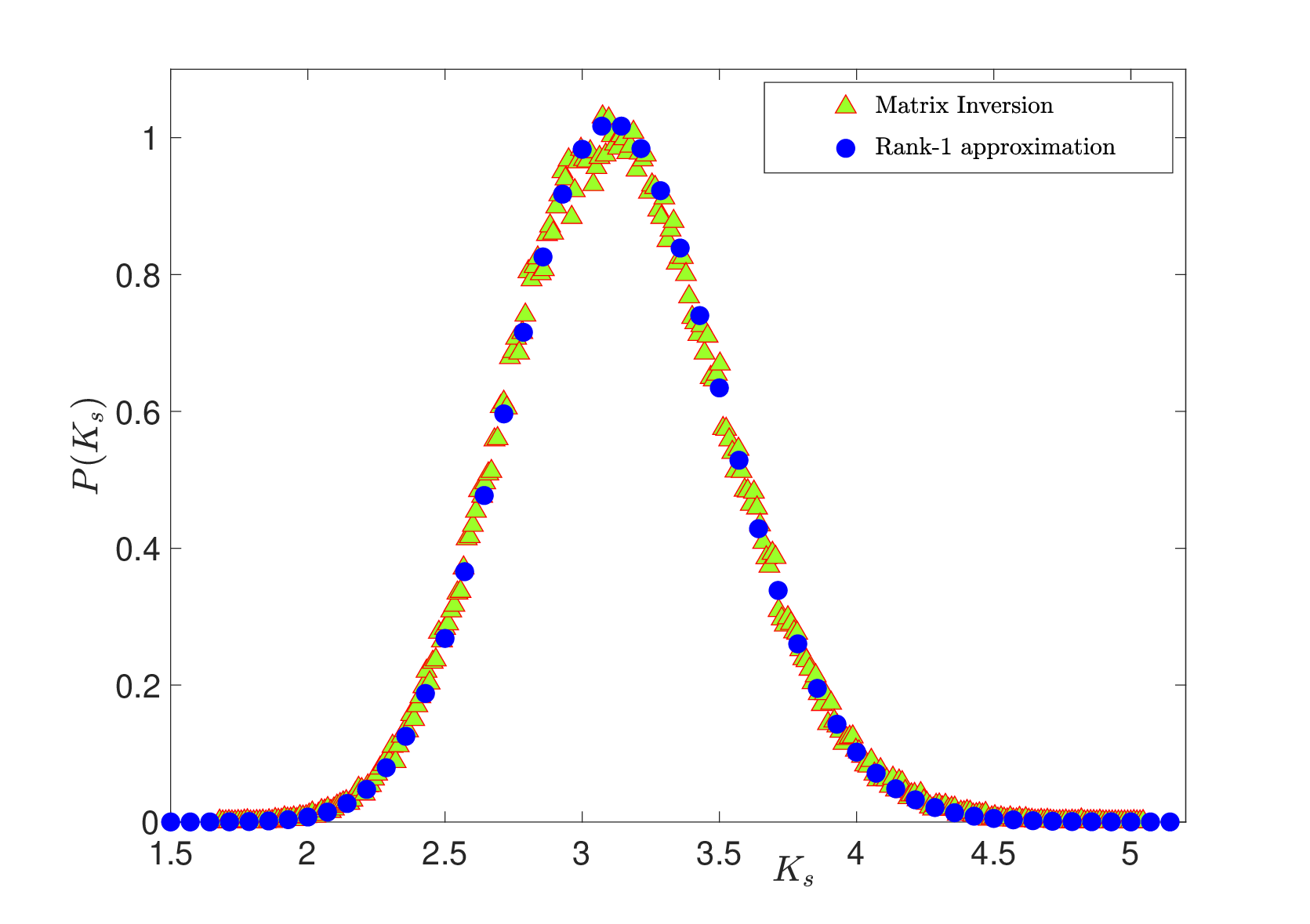}}
    \caption{Probability density function $P(K_s)$ of the shifted Katz centrality $K_s$ for an ensemble of $30$ Erd\H os-R\'enyi networks of size $N=5000$ with $c=30$ and $\alpha=1/45$. Green triangles: histogram of node centralities from randomly generated E-R networks using the inversion formula \eqref{defKatzvector}. Blue dots: Dirac comb approximate formula \eqref{eq:Diraccomb} with $K=K_s-1$.}
    \label{fig:Rank1_1}
\end{figure}

\newpage
\begin{figure}[h]
    \centering
    \fbox{\includegraphics[scale = 0.29]{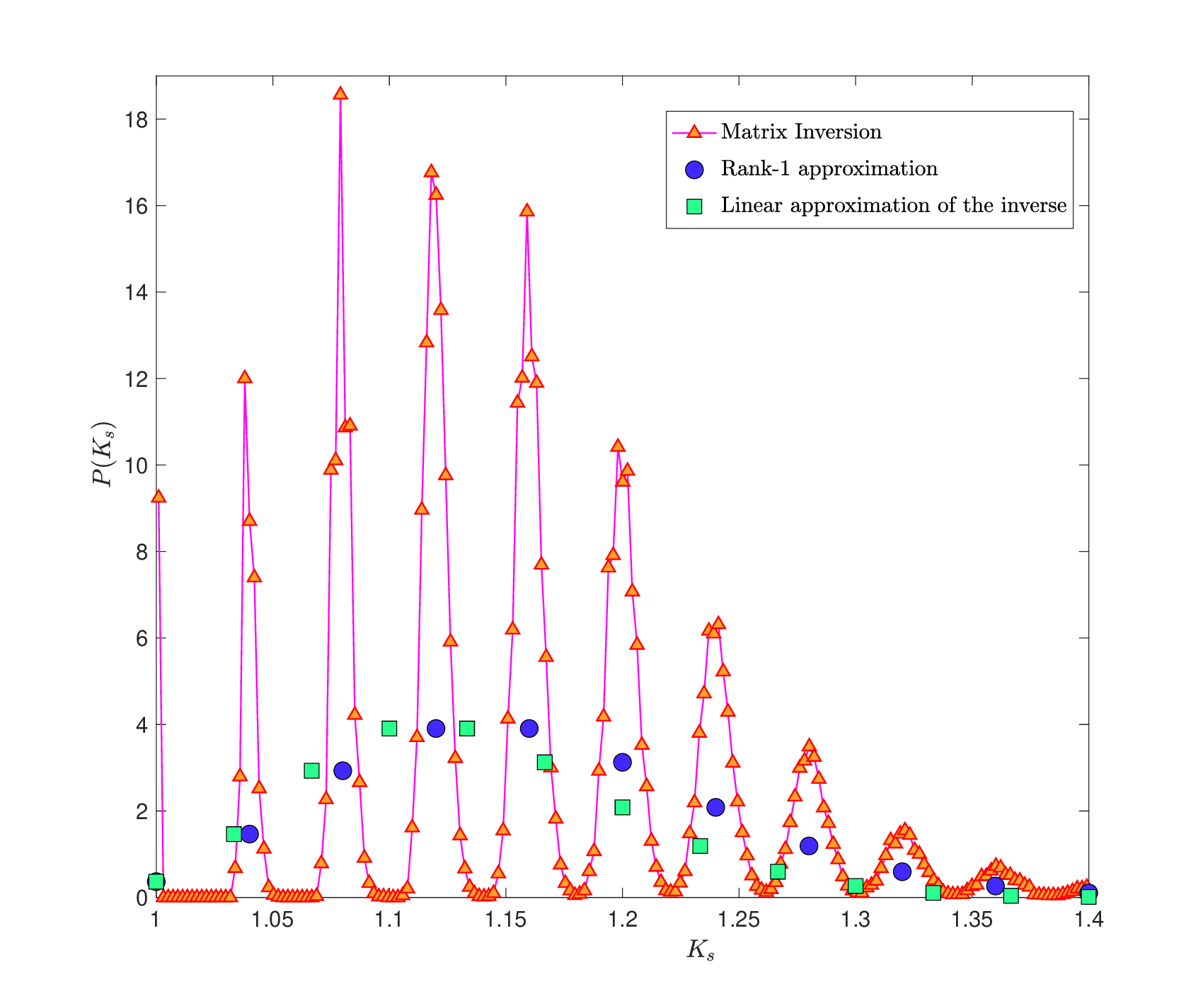}}
    \caption{Probability density function $P(K_s)$ of the shifted Katz centrality $K_s$ for an ensemble of $30$ Erd\H os-R\'enyi networks of size $N=5000$ with $c=4$ and $\alpha=1/30$. Orange triangles: histogram of node centralities from randomly generated E-R networks using the inversion formula \eqref{defKatzvector}. Blue dots: Dirac comb approximate formula \eqref{eq:Diraccomb} with $K=K_s-1$. Green squares: Dirac comb formula resulting from a simple linear approximation of the resolvent (see Eq. \eqref{eq:lineartruncresolvent}). The $y$-values of both the blue and green points have been magnified by a factor $20$ to make them visible on the same scale.}
    \label{fig:Rank1_2}
\end{figure}

\section{Conclusions and Outlook}\label{sec:concl}

In this work, we considered the distribution of the Katz centrality of nodes on single instances and on ensembles of undirected random graphs in the locally tree-like regime, focussing in particular on Erd\H os-R\'enyi and Scale Free networks. The Katz centrality of a node is a measure of how important that node is in the context of information flow across the network, as it is a weighted sum of paths of all lengths reaching that node from all other nodes, where longer paths are weighted less by a factor $\alpha$. Having accurate analytical control over the full distributions in ``null models'' (with interactions drawn at random with a prescribed distribution) is important to provide a benchmark to gauge deviations arising in empirical and synthetic data. Quite unexpectedly, though, the available analytical results are remarkably scarce, which motivates the work we presented here. 

The (shifted) Katz centralities of all nodes satisfy a linear system of equations (see \eqref{defKs}), which can be efficiently solved on a single instance of the network using the cavity method (or Gaussian Belief Propagation algorithm). We reviewed in detail the underlying theory in Section \eqref{sec:linearsys}. 

From the single instance solution, it is straightforward to deduce that the probability $P(K)$ of observing a node with centrality $K$ in an ensemble of random networks can be computed from the functional solution of a pair of recursive distributional equations (see Eqs. \eqref{eq:ch1_cavity_pi} and \eqref{eq:pitilde}), which can be efficiently solved using a Population Dynamics algorithm as described in Section \ref{sec:popdyn}.

Our results further confirm that the Katz centrality is highly correlated with the degree of nodes, with the $k$-th peak in the distribution precisely corresponding to the contributions of nodes of degree $k$ to the centrality. The sharply multimodal distribution of the centrality for low $c$ gradually crosses over towards a unimodal distribution as the average degree $c$ increases, with different peaks merging together. 

Moreover, we have provided an analytical approximation for the centrality distribution, which is based on the rank-$1$ projection proposed in \cite{bart1} and works well for not-too-sparse graphs. If the graphs are very sparse, the approximation is anyway able to capture the location and mass of each peak in a more accurate way than a simple linear truncation of the resolvent matrix. 

It will be interesting to modify the treatment presented here to deal with the case of networks with correlated degrees, as well as \emph{directed} networks for which the GaBP/cavity solution of a linear system \eqref{linearsys} on a tree structure requires some changes \cite{BPnonsymmetric}. Extending the analysis to non-symmetric adjacency matrices would allow us to deal for instance with the distribution of PageRank in random networks, a topic that has received some attention in the mathematical literature lately in the context of the so-called `power-law hypothesis' described in the Introduction.

\newcommand\ackcontent{The work of F. Caravelli was carried out under the auspices of the NNSA of the U.S. DoE at LANL under Contract No. DE-AC52-06NA25396. F. Caravelli was also financed via DOE LDRD grant 20240245ER.  P.V. acknowledges support from UKRI Future Leaders Fellowship Scheme
(No. MR/S03174X/1). For the purpose of open access, the authors have applied a Creative Commons Attribution (CC BY) license to any author-accepted manuscript version arising. We also acknowledge the stimulating environment at the Conclave on
Complexity in Physical Interacting Systems (Santa Fe, July 2023) where this work was initiated, as well as the NetSciX 2024 conference in Venice, where further substantial progress was achieved. }

\ifpnas
\showmatmethods{} 

\acknow{\ackcontent}
\showacknow{} 

\else 
\section*{Acknowledgments}
\ackcontent

\fi

\end{document}